# On fluctuation properties of the cumulative current in gTASEP on open chains

N.C. Pesheva[(a)*], N.Zh. Bunzarova[(a,c)], E. K. Mitsova[(a)], and A.M. Povolotsky[(b,d)]

*[(a)]Institute of Mechanics, Bulgarian Academy of Sciences,*
*Acad. G. Bonchev St., block 4, Sofia 1113, Bulgaria, email: nadezhda@imbm.bas.bg*
*[(b)] Bogoliubov Laboratory of Theoretical Physics,*
*Joint Institute for Nuclear Research, Dubna 141980, Moscow reg., Russia*
*[(c)] South-West University "Neofit Rilski"*
*66 Ivan Mihaylov st., 2700, Blagoevgrad, Bulgaria*
*[(d)] National Research University Higher School of Economics,*
*20 Myasnitskaya, 101000 Moscow, Russia*

**Abstract**
We present preliminary results of our numerical study of the fluctuation properties of the cumulative current, $J_T$, in the generalized totally asymmetric simple exclusion process (gTASEP) on open chains with Liggett boundary conditions. GTASEP is the ordinary TASEP (a model of 1D- traffic on a single track, having many diverse applications), where an additional term controlling clustering of the particles is added. Extensive numerical simulations are performed to determine the dependence of the second scaled cumulant $C_2$ of the time integrated current, $J_T$, on the system size, $L$, in the long-time limit, allowing for comparison between open and periodic geometries. Preliminary results for $C_2$ in the maximum current phase show agreement with theoretical predictions about the Kardar-Parizi-Zhang universality of the gTASEP on a segment.

## 1. Introduction

The generalized TASEP (gTASEP) [1-4] extends the standard model (see, e.g., [5-8] and references therein) by incorporating effective interactions between particles, representing attractive or repulsive tendencies, arising from biochemical or cooperative mechanisms. This makes it possible to study aggregation-fragmentation processes, fluctuations, finite-size effects, and boundary-induced phase behavior in nonequilibrium stationary states. Thus, TASEP and gTASEP provide useful theoretical tools for linking microscopic update rules with macroscopic transport and biological function, especially in systems governed by crowding, limited resources, and collective motion.

Biological transport is intrinsically nonequilibrium and often involves directed motion of interacting molecules along effectively one-dimensional tracks. TASEP provides a minimal stochastic description of such processes [5,6], capturing molecular crowding, boundary-controlled flux, and collective effects. Originally introduced in 1968 as a model for protein synthesis kinetics [9], it has been widely applied to ribosome motion along messenger RNA during translation and to molecular motor proteins, such as kinesin and dynein, moving along microtubules [8]. These mechanisms are central to intracellular transport in eukaryotic cells.

Beyond applications to transport phenomena, exclusion processes play an important role in the investigation of universal fluctuation properties. In particular, TASEP is a prominent microscopic representative of the Kardar-Parisi-Zhang (KPZ) universality class [10, 11]. Initially formulated to describe stochastic interface growth, KPZ universality also encompasses driven particle systems, directed polymers in random environments, certain problems in combinatorics and random matrix theory [11, 12] etc.. These models exhibit related large-scale fluctuation statistics despite differences in their microscopic rules.

Models belonging to the KPZ universality class share several essential features [10,11], including stochastic forcing, typically represented by short-range-correlated or white noise, local dynamics, in which the evolution of the system depends on its immediate environment, and a nonlinear contribution associated with the local slope of the growing interface. The interplay between stochastic fluctuations, nonlinear dynamics, and spatial smoothing gives rise to the characteristic scaling properties of the KPZ universality class.

For exclusion processes, a height-function mapping identifies particle jumps with local increments of a growing interface. As a result, fluctuations of the time-integrated particle current can be studied using tools developed for KPZ growth [6, 11]. This connection motivates the investigation of current cumulants and their dependence on observation time, system size, and boundary conditions.

Although gTASEP has been extensively studied on periodic and infinite lattices [2-4,13], its behavior on finite open chains introduces additional questions, particularly because of the influence of particle reservoirs at the boundaries. Previous studies of aggregation–fragmentation dynamics on open chains [4] and the analytical phase diagram obtained for gTASEP with Liggett-like boundary conditions [14] provide a foundation for addressing these questions. Building on these results, the present work goes beyond the analysis of stationary density profiles and mean particle currents to investigate fluctuations of the time-integrated particle current and their dependence on system size, with particular attention to the maximal-current phase

## 2.Description of the model system

We study the non-equilibrium steady states (NESS) of the generalized totally asymmetric simple exclusion process on open lattices with Liggett boundary conditions (BCs). The model describes a system of hard-core particles moving along a one-dimensional (1D) lattice governed by stochastic, discrete-time, and discrete-space kinetics under the simple exclusion principle (see Fig. 1).

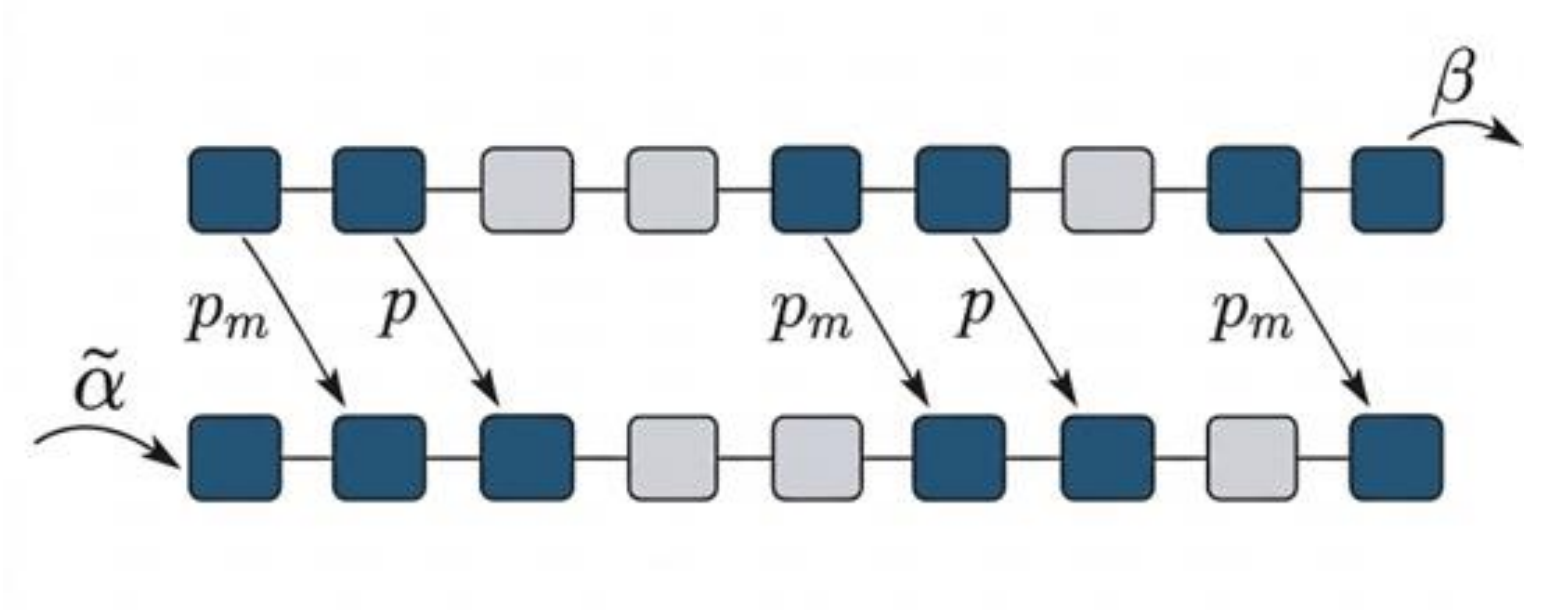

Fig. 1. Schematic diagram of the generalized TASEP on open chains and input-output rates ($\alpha$, $\beta$). A second hopping probability, $p_m$ ($p_m \equiv \mu$ in the notation, used in Ref.[14]), is introduced in the model for particles within a cluster, in addition to the standard hopping probability, $p$ in ordinary TASEP. The update rules and the Ligget BCs for the temporal evolution of the system are explained in detail in the text (in some cases $\tilde{\alpha}$ (Eq. 1) is used instead of $\alpha$). Here, they are illustrated by showing possible configurational changes through two consecutive time steps. This example represents the case when $p < p_m < 1$, favoring clustering of particles (i.e., attractive interaction between particles in a cluster).

More, precisely, each site $i$, $i = 1,2,\dots,L$ can hold at most one particle $n_i = 0, 1$ (where $L$ is the size of the system). Fundamentally, this model extends the standard TASEP with backward sequential update (BSU). Thus, in addition to the standard forward hopping probability $p$, a modified hopping probability $p_m$ is introduced specifically for particles within a cluster. This additional parameter models a kinematic interaction between adjacent particles belonging to the same cluster. The time evolution of the system obeys the following rules: The update sweeps consecutively through the whole chain, starting from the last site $i = L$ (i.e., BSU is used). Particles are ejected from the last site of the chain, $i = L$, with probability $\beta$. In the bulk of the system: a single particle or the lead particle of a cluster can jump on an empty site to the right with probability $p$ , the particles in the cluster may jump to the neighboring site to the right with probability $p_m$, provided the particle in front of it has already jumped. The site, $i=1$, is the last to be updated: If the site $i = 1$ is empty at the beginning of the current update, a particle enters the system with probability $\alpha$, or site $i = 1$ remains empty with probability $1 - \alpha$; If the site, $i = 1$, is occupied at the beginning of the current moment of time, but became empty under its current update, then a particle enters the system with a modified probability (Ligget BCs) [14]:

(1) $$\tilde{\alpha} = \frac{\alpha\, p_m(1-p)}{\alpha(p_m-p)+p(1-p_m)}\,.$$

The model is viewed as a generalization of the standard TASEP, since, for the specific values of $p_m = 0$ and $p_m = p,$ it reduces to the well known cases of TASEP with parallel update (PU) and TASEP with BSU. The Ligget BCs model the system being attached to the particle reservoirs, where the particle jumps have the same probabilities as in the stationary infinite translation invariant gTASEP.

## 3. Numerical results and discussion

The aim of the present work is to explore the fluctuation properties of the cumulative current $J_T$ in gTASEP on open chains with Liggett BCs. The goal is to find out whether KPZ universality is observed in the current model. The KPZ- universality in the gTASEP model is expected to manifests itself in the asymptotic behavior (at long times) of the scaled cumulants $C_n$ of the quantity $J_T$, which represents the number of particles that have entered or left the system by time T in the system's non-equilibrium steady state. In other words, $J_T$ is the cumulative current in the system consisting of $L$ sites. Thus, $J_T$ is defined as:

(2) $$J_T=\langle \textstyle\sum_{t=1}^{T} j_{in}(t)\rangle\ ,$$

where $j_{in}(t)$ (respectively, $j_{out}$ (t)) is the incoming/ outcoming current at time moment *t.* The scaled cumulants $C_n$ are defined as the limit when $t \to \infty$ of the relations:

(3) $$C_n = \lim_{t\to\infty} \frac{\langle J_t^n\rangle_c}{t} \qquad n=1,2,$$

The standard cumulants $\langle J_t^n\rangle_c$ (the subscript *c* stands for cumulant) are defined as:

(4) $$\langle J_t^n\rangle_c = \frac{d^n}{ds^n}|_{s=0} \ln \langle e^{sJ_t^n}\rangle\,,$$

where $\langle e^{sJ_t^n}\rangle$ is the moment generating function (MGF) for the physical quantity $J_t$, *s* is a dummy variable. In particular,

(4.1) $n=1$ $\quad \langle J_t\rangle_c=\langle J_t\rangle\ ;$

(4.2) $n=2$ $\quad \langle J_t^2\rangle_c = \langle J_t^2\rangle - \langle J_t\rangle^2.$

As usual the angular brackets ⟨ ⟩, denote the averaging of the respective quantity over the ensemble of $N_{ens}$ systems in NESS.

The numerical results presented below (see Figs. 2 and 3) characterize the generalized TASEP, featuring attractive particle interactions. Specifically, we investigate the parameter regime, where the standard hopping probability is $p$=0.75 and the cluster hopping probability is $p_m$=0.9. The particle injection and ejection rates are set symmetrically to $\alpha\ (\equiv P_{in}) = 0.85$ and $\beta\ (\equiv P_{out}) = 0.85$, respectively. As established in prior studies [14], this specific parameter set positions the system deep within the MC phase.

### *3.1. Temporal evolution and steady-state relaxation*

We first examine the long-time temporal dependence (Fig. 2a) of the scaled current cumulants (Eq. (3)). Similarly to the bulk density, $\rho$, defined as:

(5) $$\rho = \langle \tfrac{1}{L}\textstyle\sum_{i=1}^{L} n_i\rangle,$$

the second scaled cumulant of the cumulative current $C_2$ eventually reaches a plateau, corresponding to the non-equilibrium steady state, where it fluctuates around a constant mean value regardless of the initial configuration. However, our numerical tests indicate that calculating $C_2$ requires significantly longer time sequences and more extensive ensemble averaging than standard bulk observables like density $\rho$ or stationary current $j$.

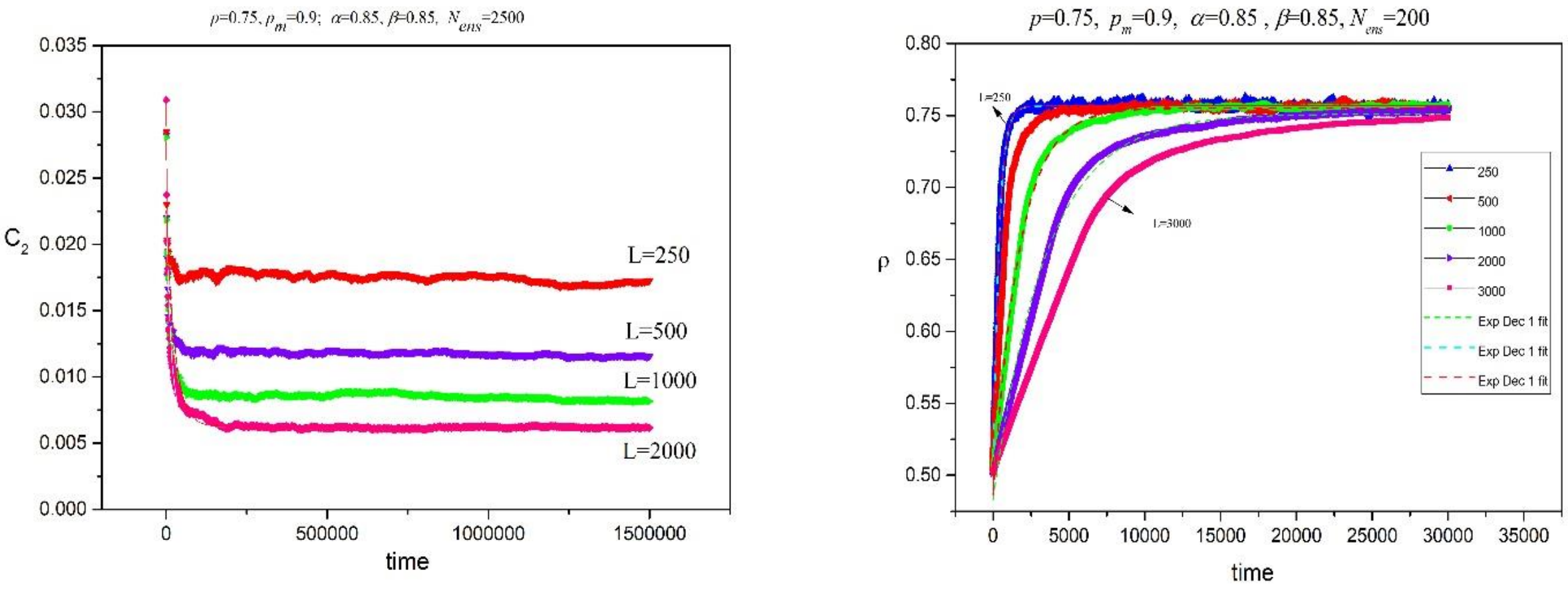

a) b)

Fig. 2. Time dependence a) of the scaled cumulant, $C_2$, of the cumulative current ($J_T$) and b) of the bulk density, $\rho$, for various system sizes: $L$ = 250, 500, 1000, 2000 sites. Time is measured in MC steps. The presented results correspond to $p = 0.75$, and $p_m$=0.9. The particle injection probability is $\alpha = 0.85$, and the particle exit probability is $\beta = 0.85$ (i.e., the simulation data are obtained at a point within the maximum current phase); in a) $N_{ens}$=1000, and in b) $N_{ens}$ =200. All simulations are started from a half filled lattice $p_I$=0.5.

As illustrated in Fig. 2(a), the relaxation time, required for $C_2$ to reach the NESS, is significantly larger than that of the bulk density. This relaxation time depends strongly on the system size, for instance, a lattice of $L$= 2000 sites requires approximately 25 000 MC time steps to reach steady state, while the relaxation time of the bulk density $\rho$ is approximately 3500 MC steps for the same set of parameters and the same system size. The data, presented in Fig. 2(a), were obtained using an ensemble size of $N_{ens}$=2500, total time sequences of $N_T = 1{,}5\text{x}10^6$ MC steps, sampled every $N_{step}$=2000 MC steps.

### *3.2. Finite-size scaling and universality class*

From the steady-state plateaus in Fig. 2(a), one can extract the long-time values of $C_2$ across various system sizes $L$. The resulting finite-size scaling of $C_2$ within the maximal-current phase is plotted in Fig. 3. To evaluate the scaling behavior, the data were fitted using three candidate functions (represented by dashed and dashed –dotted lines in Fig. 3):

1. An exponential decay function;
2. A free power law, $f(x)=a\,x^b$ , where both $a$ and $b$ are free parameters;
3. A fixed power law, $f(x)=a\,x^{-0.5}$ with the exponent constrained to $b$=-0.5.

Notably, enforcing the fixed exponent $b$=-0.5, yields an even lower reduced $\chi^2$ value than the free power-law fit, while the adjusted $R^2$ remains virtually unchanged. Single exponential decay function produces the worst fit.

This scaling exponent of $b$ = -0.5 is highly characteristic of models, belonging to the KPZ universality class. Furthermore, these preliminary results for the prefactor $a$ show excellent agreement with existing theoretical predictions [15] for the universal behavior of $C_2$ in the maximal-current phase. To definitively confirm that the fluctuation properties of the gTASEP on open chains, governed by Liggett BCs, fall into the KPZ universality class, more extensive Monte Carlo simulations and rigorous analytical calculations are currently underway.

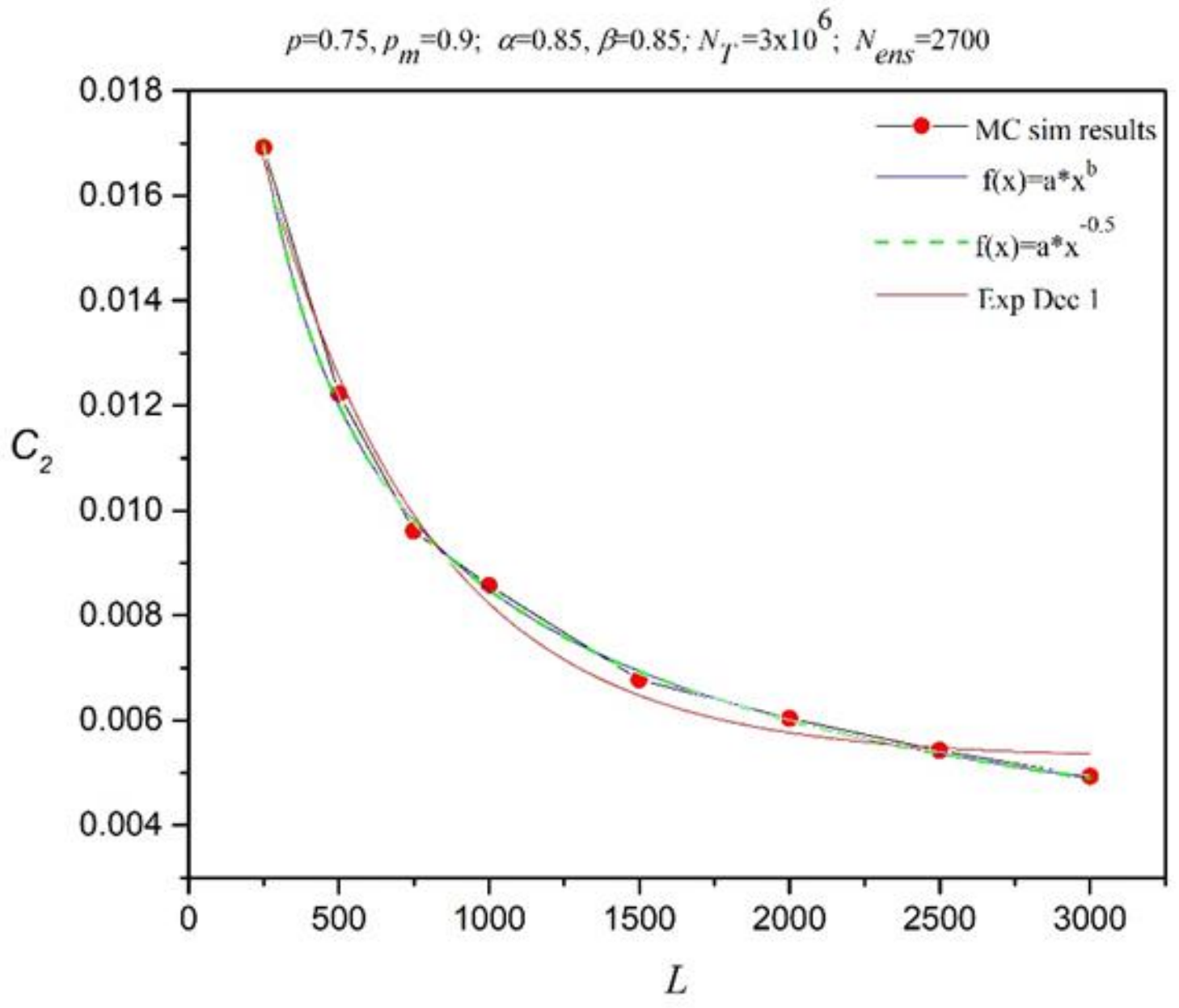

Fig. 3. Dependence of the second cumulant $C_2$ (2) of the cumulative current ($J_T$ ) on the system size $L$ (in the interval $L \in [250,3000]$ sites). The presented results are for $p = 0.75$, and $p_m$=0.9, the probability of a particle entering the system is $\alpha$=0.85, and the probability of a particle leaving the system is $\beta$=0.85, i.e., the simulation is performed at a point belonging to the maximum-current phase. The ensemble averaging is $N_{ens}$=2 700, the length of the time sequences is $N_T = 3\text{x}10^6$ MC steps, and $N_{tau}$=200000 MCs are omitted to ensure that the system has reached a NESS.

## 4. Conclusion

Using numerical simulations, we investigate the second cumulant, $C_2$, of the time-integrated current, $J_T$, and its dependence on the system size $L$ in the long-time regime. Particular attention is paid to the maximal-current phase, where collective fluctuations play an important role. Our primary finding is that the scaling of $C_2$ with $L$ in the long time limit is a well-characterized by a power law with an exponent of $b$=-0.5. This exponent is a hallmark of models within the Kardar–Parisi–Zhang universality class and is consistent with theoretical predictions for the universality of current fluctuations [15]. However, more extensive numerical simulations and ongoing theoretical calculations are required to definitively confirm that the fluctuation properties of the gTASEP on chains with Liggett BCs indeed belongs to the KPZ universality class.

## Acknowledgements

N.P. would like to thank for partial support the Centre of Excellence in Informatics and ICT under the Grant No BG16RFPR002-1.014-0018, financed by the Research, Innovation and Digitalization for Smart Transformation Programme 2021-2027 and co-financed by the European Union.
N.B. and N.P. gratefully acknowledge the hospitality of the Bogoliubov Laboratory of Theoretical Physics, JINR, Dubna.